\documentclass[aps,twocolumn,pra,superscriptaddress,showpacs,tightenlines]{revtex4}

\usepackage{amssymb}
\usepackage{amsmath}
\usepackage{graphicx}
\usepackage{epsfig}
\usepackage{subfigure}
\usepackage{amsfonts}

\begin{document}

\title{Transferring and bounding single photon in waveguide controlled \\
by quantum node based on atomic ensemble}
\author{Jing Lu}
\affiliation{Key Laboratory of Low-Dimensional Quantum Structures
and Quantum Control of Ministry of Education, and Department of
Physics, Hunan Normal University, Changsha 410081, China}
\affiliation{Institute of Theoretical Physics, Chinese Academy of
Sciences, Beijing, 100080, China}
\author{H. Dong}
\affiliation{Institute of Theoretical Physics, Chinese Academy of Sciences, Beijing, 100080, China}
\author{Le-Man Kuang}
\affiliation{Key Laboratory of Low-Dimensional Quantum Structures
and Quantum Control of Ministry of Education, and Department of
Physics, Hunan Normal University, Changsha 410081, China}

\begin{abstract}
We study the scattering process of photons confined in a one dimensional
optical waveguide by a laser controlled atomic ensemble. The investigation
leads to an alternative setup of quantum node controlling the coherent transfer
of single photon in such one dimensional continuum. To exactly solve the
effective scattering equations by using the discrete coordinate approach,
we simulate the linear waveguide as a coupled resonator array at the high
energy limit. We generally calculate the transmission coefficients and its
vanishing at resonace reflects the good controllability of our scheme.
We also show that there exist two bound states to describe the
localize photons around the cavity.
\end{abstract}

\maketitle

\section{\label{Sec:introduction}Introduction}

Single photon source and single photon detection is very crucial to the
quantum information processing\cite{Sun1}. Essentially, this importance is
\textquotedblleft rooted in\textquotedblright \
the coherent manipulation of single photon through the controllable quantum
node at the single photon level. There are many protocols to implement
such quantum node, for example, a two or three-level system
(hereafter the term \textquotedblleft atom\textquotedblright \
is used to refer to these kind of systems) inside
a waveguide\cite{Lukin-np,fanL95,dong} based on the absorption and
reemission of the single photon. However, the absorbing cross section is
much smaller in this kind of quantum node. When an atom is inserted into a
cavity with high quality factor, the cross section will be largely improved
between an atom and single photon due to photon moving to and fro inside the
cavity. With this microcavity based setup, quantum nodes are also proposed
on coupled-resonator array by embedding a two or three level
atom\cite{ZGLSN,gongzr}. Even with such cavity enhanced coupling a much
stronger coupling is hardly to achieve. An alternative approach to alter
this situation is to interact photons with a coherent ensemble of
atoms\cite{sunl91,LiyA69,Lukin84,mLukin84,Phillip86}.
Here, the atomic coherence results from the fact the
inter-atomic distance is less than the wavelength
of their radiation from atoms. Therefore, if one naturally replace the
single atom with an ensemble of N identical atoms, a $\sqrt{N}$-time
enhancement appears in the effective coupling.

Actually, this enhancement effects have been extensively explored for the
free space case or in a single cavity\cite{Lukin-rmp03,sunyiyou-pra03}.
In this paper, we will concern this
ensemble enhancement effect in the transferring photons confined in
waveguide. We understand this case as a semi-free case where all atoms are
confined in a cavity while the photon can propagate in the waveguide (see
the Figure \ref{qnode-1}).
In present investigation we adapt the discrete coordinate
approach of scattering\cite{ZGLSN,gongzr}. In
this approach the waveguide is approximated as a coupled resonator array
with some dispersion relations.
The physics of such approximation is rooted in the studies of photonic
crystal waveguide defect based\cite{PCrystals}. This method seems to be
universal since it can give the exact solutions both in the continuum limits
with higher and lower energy\cite{ZGLSN}.
We first model the 1D waveguide as
a coupled resonator array and also consider the collective excitation of an
atomic ensemble in the single excitation with lower energy. Then, the photon
transfer in the waveguide is described as a coherent hopping of photon along
the coupled cavity array. The atomic ensemble localized in a cavity within
this array behaves as the scatterer. By using of the
scattering theory, we obtain the single photon transmission rate in the
coupled resonator array. And we find there exist two bound states to
describe the localize photons.

This paper is organized as follow: in Sec.\ref{Sec:model}, we present our
model, a coupled-cavity array with atomic ensemble separately inside
its corresponding cavity. In Sec.\ref{Sec:a-exci}, we consider the transport
property in one-excitation subspace under discrete coordinate approximation.
Then we consider the role of the enhanced coupling for the
single photon transfer in the coupled resonator array in Sec.\ref{Sec:exact-s}.
Through the scattering equation in the discrete coordinate representation,
we study the photon scattering by the atomic ensemble in Sec.\ref{Sec:control}.
In Sec.\ref{Sec:boundstate}, we also obtain the two bound states
to describe the photon localization.
Then we make our conclusions in Sec.\ref{Sec:conclusion}.

\section{\label{Sec:model}Modeling hybrid system for controlling photon transfer}

\begin{figure}[tbp]
\includegraphics[bb=62 16 382 260, width=8 cm,clip]{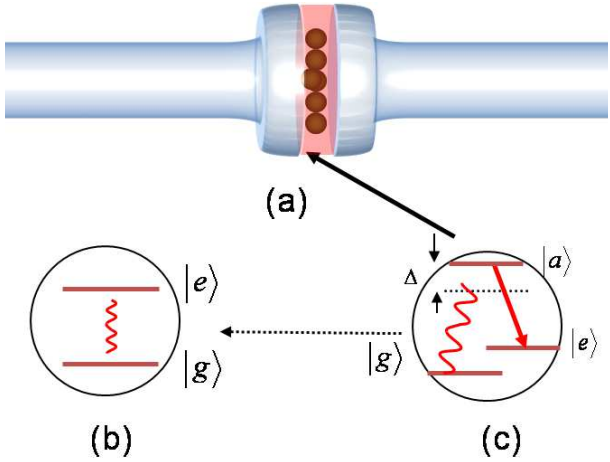}
\caption{(Color online) Schematic map of quantum node for single photon control
(a) One waveguide coupled with an atomic gas cell.
(b) The gas cell is full of two-level atoms. (c) To implement the
steady two-level atoms, the stimulated Ramman process based on $\Lambda$%
-type atoms is used to achieve the effective two-level atoms in large
detuning to overcome the high-level decay.}
\label{qnode-1}
\end{figure}

Schematic description of our system is shown in Fig.\ref{qnode-1}
where photons propagating in the one dimensional waveguide are
coupled to the atomic ensemble in a gas cell.
We assume there is no inter-atom interaction for a dilute gas in the cell.
The characteristic size of the gas cell is comparable to the wave length
of the atomic radiation. The waveguide supports the input and
output propagating optical modes.
To put the our system into a mathematical formalism, we
introduce the bosonic field operator $\varphi _{R}=\varphi _{R}\left(
x\right) $ and $\varphi _{L}=\varphi _{L}\left( x\right) $ for a right-going
and a left-going photon respectively\cite{fanL95}, then its corresponding Hamiltonian
reads%
\begin{equation}
H_{w}=-iv_{g}\int_{0}^{\infty }dx\left( \varphi _{R}^{\dag }\partial
_{x}\varphi _{R}-\varphi _{L}^{\dag }\partial _{x}\varphi _{L}\right) \text{,%
}  \label{hwrh-01}
\end{equation}
where $v_{g}$ stands for the group velocity of the photon in the waveguide.

The gas-cell, localized around point $x=0$, is filled in with an ensemble of
two level systems. And this ensemble is coupled to input and output fields
inside waveguide. Actually, in order to achieve a tunable two-level atomic
ensemble, we employ $N$ three-level atoms of a $\Lambda $-type level
structure: the ground state $\left\vert g\right\rangle $, the excited state $%
\left\vert e\right\rangle $ and an auxiliary state $\left\vert
a\right\rangle $, shown in Fig. \ref{qnode-1}(c). The transition between
levels $\left\vert e\right\rangle $ and $\left\vert a\right\rangle $ is
driven by a classical control field, and the transition between levels $%
\left\vert g\right\rangle $ and $\left\vert e\right\rangle $ couples via
dipole moments to the cavity resonance mode. Through the stimulated Raman
process, a tunable and much stable two-level atomic ensemble is
achieved\cite{HFMA76}.

The model Hamiltonian for the total system reads
\begin{equation}
H^{h}_{JC}=\Omega \sum_{l=1}^{N}\sigma _{ee}^{l}+\sum_{l=1}^{N}\xi\left[
\zeta_{l}\left( \varphi _{L}^{\dag }\left( 0\right) +\varphi _{R}^{\dag }\left(
0\right) \right) \sigma _{ge}^{l}+\mathrm{h.c.}\right] \text{,}
\label{hwrh-02}
\end{equation}%
where $\sigma _{\mu \nu }^{l}=\left\vert \mu \right\rangle _{l}\left\langle
\nu \right\vert ,\left( \mu ,\nu =e,g\right) $ flips the energy level $\left\vert
\nu \right\rangle $ of atom $l$ to $\left\vert \mu \right\rangle $ of
the same atom, $\Omega $ is the energy level spacing of the atomic system,
$\xi$ is the light-atom coupling strength in the waveguide,  and $\zeta_{l}$
$(\zeta_{l}<1)$ characterizes the inhomogeneity of couplings of each atom
to the cavity photons. In experiments, the coupling coefficients $\xi$ depends on
the position of the atom, but we take it uniform for an idealized
consideration to abstract the dominate conclusions.

Next, we can deal with the photon transfer in the waveguide as the
photon hopping in an infinite array of coupled cavities and
central cavity contains an identical atomic ensemble.
\begin{figure}[tbp]
\includegraphics[bb=93 498 474 621, width=7 cm,clip]{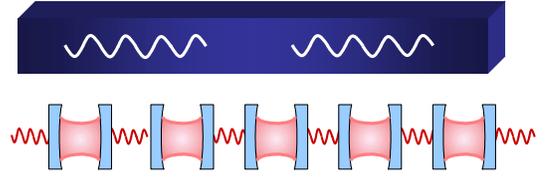}
\caption{(Color online) Simulation of the waveguide by a coupled resonator
array for the \textquotedblleft high energy \textquotedblright photon hoping inside.}
\label{qnode-2}
\end{figure}
In the physical implementation, such kind of one-dimensional coupled
resonator waveguide (CRW) can be realized by coupled superconducting
transmission line resonators~\cite{zhgsun}, coupled photonic-crystal
microcavities or fibre-coupled toroidal microcavities. We will prove
as follows that, for the \textquotedblleft high energy\textquotedblright \
photons, the coupled resonator array will play the same role as that
by the waveguide.

Usually, such nearest-neighbor interactions of cavities are modeled as a
typical tight-binding coupling in terms of the creation and annihilation
operators ($b_{j}^{\dag }$ and $b_{j}$) of the modes localized in the $j$th
cavity. The model Hamiltonian for the CRW yields
\begin{equation}
H_{C}=\sum_{j}\omega b_{j}^{\dagger }b_{j}-\sum_{j}g\left( b_{j}^{\dagger
}b_{j+1}+b_{j+1}^{\dagger }b_{j}\right) \text{,}  \label{qndis-01}
\end{equation}%
where $\omega $ is the eigenfrequency of the cavity mode, and
$g$ is the hopping constant between the neighboring sites. Here,
the subscript index $j$ range from minus infinite to plus infinite.

The photonic spectrum of this CRW is continuous with its band extending from $\omega
-2g$ to $\omega +2g$. A propagating single photon in the CRW will occupy an
energy
\begin{equation}
E_{k}=\omega -2g\cos k  \label{qndis-02}
\end{equation}%
where $k$ is the momentum of the photon, and the lattice spacing is taken to
be unit. In the low-energy regime $k\rightarrow 0$, the long-wavelength
approximation gives a quadratic spectrum%
\begin{equation}
E_{k}\simeq E_{k}^{l}\equiv \omega _{g}+gk^{2}\text{,}  \label{qndis-03}
\end{equation}%
with $\omega _{g}=\omega -2g$, which is obtained by expanding the cosine
function around zero. Then this system is reduced to the one that a particle
with mass $\left( 2g\right) ^{-1}$ moves in a free space or a waveguide with
no energy bound.

In the high-energy regime $k\rightarrow \pi /2$, the
short-wavelength approximation leads to a linear spectrum
\begin{equation}
E_{k}\simeq E_{k}^{h}\equiv \omega _{\pi }\pm 2gk\text{,}  \label{qndis-04}
\end{equation}%
with $\omega _{\pi }=\omega -\pi g$, which is obtained by expanding the
cosine function around $\pm \pi $.

\section{\label{Sec:a-exci}Quasi-spin wave excitations of atomic ensemble
coupled to the photon hopping}

In this section, we model the interaction between the hopping photons and the
collective excitations of the atomic ensemble. Similar to the usual spin wave
in the magnetic system, this collective excitation is described by a collective
operator
\begin{equation}
a^{\dag }=\frac{1}{\sqrt{N(\zeta)}}\sum_{l=1}^{N}\zeta_{l}\sigma _{eg}^{l}\text{,}
\label{qndis-05}
\end{equation}
where $N(\zeta)=\sum_{l=1}^{N}|\zeta_{l}|^{2}$. Obviously, the above collective
operators describe the collective excitation from the ground state $\left\vert
G\right\rangle =\left\vert g_{1}\cdots g_{N}\right\rangle $ with all atoms
in the ground state. A single particle excitation is presented by
$\left\vert 1_{a}\right\rangle=a^{\dag}\left\vert G\right\rangle $.
In the large $N$ limit, and under the low excitation
condition that there are only a few atoms occupying the excited state
$\left\vert e\right\rangle $, the collective operator $a$ satisfy the
commutation relation
\begin{equation*}
\left[ a,a^{\dag }\right] \xrightarrow{N\rightarrow\infty}1\text{.}
\end{equation*}
This means that the quasi-spin wave excitation is of bosonic type\cite{JGRSCP}.

According to the reference \cite{song-prb05}, there also exist other collective
modes, but they are decoupled with this mode by $a^{\dag}$ when the atoms is in the
homogeneously broadening, i.e., all the energy level spacings are the same.
Then the coupling between the atomic ensemble
and the localized mode $b_{0}$ and $b_{0}^{\dagger }$ in 0'th cavity reads
\begin{eqnarray}
H_{JC} &=& \Omega a^{\dagger }a+\sqrt{N}\xi \left( b_{0}^{\dagger }a+a^{\dagger
}b_{0}\right)   \label{qndis-06}
\end{eqnarray}%
where $\xi $ the cavity mediated atom-photon coupling.

In terms of the quasi-spin-wave excitation of the atomic ensemble, the
Hamiltonian in the low-energy regime reduces to
\begin{eqnarray}
H_{l} &=&-g\int_{0}^{\infty }dx\varphi ^{\dag }\partial _{x}^{2}\varphi +\Omega
a^{\dagger }a  \notag \\
&&+\sqrt{N}\xi \left[ \varphi ^{\dag }\left( 0\right) a+a^{\dagger }\varphi \left(
0\right) \right]  \label{qndis-07}
\end{eqnarray}%
In the high-energy regime, the Hamiltonian yields
\begin{eqnarray}
H_{h} &=&-i2g\int_{0}^{\infty }dx\left( \varphi _{R}^{\dag }\partial _{x}\varphi
_{R}-\varphi _{L}^{\dag }\partial _{x}\varphi _{L}\right) +\Omega a^{\dagger
}a  \notag \\
&&+\sqrt{N}\xi \left[ \left( \varphi _{L}^{\dag }\left( 0\right) +\varphi _{R}^{\dag
}\left( 0\right) \right) a+\mathrm{h.c.}\right]  \label{qndis-08}
\end{eqnarray}%
In the above equations, we have neglected constants terms, and
the summation of $l$ over the atoms is inside a small but macroscopic volume around the
position $x=0$. Obviously, due to the
cooperative motion of all atoms, the atom-photon interaction is enhanced
by $\sqrt{N}$ times. Thus our quantum node has a strong coupling to light,
so that it can preform a fast control for the photon transfer.

\section{\label{Sec:exact-s}Coherent scattering of photons by atomic ensemble}

In this section, we consider the role of the enhanced coupling for the
single photon transfer in the coupled resonator array.
The strong coupling between the quasi-spin-wave and the
cavity field leads to the emergence of dressed states of atoms
by photons, which is called polaritons\cite{HLSCP07}.
Polaritons are quasiparticles, which were introduced to reveal the physical
mechanism for the temporary transfer of excitations to and from the medium
many years ago\cite{Dutra97,Hopfield58}.

In our system, there are two polaritons, which are described by polariton
operators
\begin{subequations}
\label{tr-01}
\begin{eqnarray}
A &=&a\cos \theta +b_{0}\sin \theta \text{,} \\
B &=&a\sin \theta -b_{0}\cos \theta \text{.}
\end{eqnarray}
\end{subequations}
Here, the superpositions of the two operators $A$ and $B$ are defined by
the angle
\begin{equation}
\theta =\arctan\left(\sqrt{\frac{\Delta -\delta }{\Delta +\delta }}\right)
\label{tr-02}
\end{equation}
with the parameters
\begin{equation}
\Delta =\sqrt{\delta ^{2}+4N\xi ^{2}}\text{, }\delta =\Omega -\omega
\end{equation}
Obviously, polariton operators $A$ and $B$ still obey bosonic commutation relations.
\begin{figure}[tbp]
\includegraphics[bb=81 526 498 692, width=7 cm,clip]{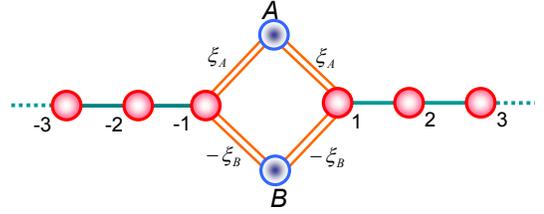}
\caption{(Color online) Equivalent configuration to Fig. \ref{qnode-2}.}
\label{qnode-3}
\end{figure}
With quasi-particle operators $A$ and $B$, the interaction between
the atoms and the $0$th cavity is
diagonalized as%
\begin{eqnarray}
H_{0} &=&\Omega a^{\dagger }a+\omega b_{0}^{\dagger }b_{0}+\sqrt{N}\xi \left(
b_{0}^{\dagger }a+h.c.\right)  \label{tr-05} \\
&=&\Omega _{+}A^{\dag }A+\Omega _{-}B^{\dag }B  \notag
\end{eqnarray}
where,
\begin{equation}
\Omega _{\pm }=\frac{1}{2}\left( \Omega +\omega \pm \Delta \right)
\label{tr-04}
\end{equation}
means that, in the atomic medium, the electromagnetic field will be split
into two displacement vectors, which depicted by $A$ and $B$
respectively, corresponding to higher energy $\Omega _{+}$ and lower energy
$\Omega _{-}$.

In terms of the two dressed collective excitation operators $A$ and $B$,
we rewrite the total Hamiltonian
\begin{align}
H& =H_{0}+\sum\limits_{j\neq 0}\omega b_{j}^{\dagger }b_{j}+\sum_{j\neq
-1,0}g\left( b_{j}^{\dagger }b_{j+1}+h.c.\right)  \notag \\
& +\xi _{A}\left( b_{-1}^{\dagger }+b_{1}^{\dagger }\right) A-\xi _{B}\left(
b_{-1}^{\dagger }+b_{1}^{\dagger }\right) B+h.c. \label{tr-06}
\end{align}
where $\xi _{A}=g\sin \theta $ and $\xi _{B}=g\cos \theta $, which depends on
the hopping constant $g$ respectively and represent the effective couplings of
$A$-polariton and $B$-polariton to the hopping photons.
This Hamiltonian describes a local two channel scattering process illustrated
in Fig. \ref{qnode-3} schematically.
The energy difference $\Delta$ between the
two channel is determined by both the Jaynes-Cummings coupling constant $\xi$
and the numbers $N$ of atom in the unit volume, and so there exists an obvious
$\sqrt{N}$-enhanced effects in the polariton split.

We confine ourselves to the single excitation subspace, since the total
excitation number $N=\sum_{j}b_{j}^{\dag }b_{j}+a^{\dag }a$ commutes with
the Hamiltonian of the system. In the coordinator representation, the
eigenstates in the single-excited subspace reads%
\begin{equation}
\left\vert \psi \right\rangle =\sum_{j\neq 0}u_{j}b_{j}^{\dagger }\left\vert
0\right\rangle +u_{A}A^{\dagger }\left\vert 0\right\rangle +u_{B}B^{\dagger
}\left\vert 0\right\rangle  \label{tr-07}
\end{equation}%
where $u_{j}$ is the probability amplitude for finding the single photon at
the $j$th site, $u_{l}$ ($l=A,B$) is the probability amplitude of the single
photon to localize at the $0$th cavity by the formation of polaritons. The
discrete Schrodinger equation gives the continuity of the wave functions in
the scattering region around the atomic gas cell
\begin{subequations}
\label{tr-09}
\begin{align}
\left( E_{k}-\omega \right) u_{-1}& =gu_{-2}+\xi _{A}u_{A}-\xi _{B}u_{B}, \\
\left( E_{k}-\omega \right) u_{1}& =gu_{2}+\xi _{A}u_{A}-\xi _{B}u_{B}, \\
\left( E_{k}-\Omega _{+}\right) u_{A}& =\xi _{A}\left( u_{-1}+u_{1}\right) ,
\\
\left( E_{k}-\Omega _{-}\right) u_{B}& =-\xi _{B}\left( u_{-1}+u_{1}\right) .
\end{align}

\section{\label{Sec:control}Photon transmission controlled by atomic ensemble}

In this section, we study the photon scattering by the two polaritons
through the above scattering equation in the discrete coordinate representation.
Actually, the above equations (\ref{tr-09}) are the re-expression of the
original discrete coordinate scattering equation first presented in Ref.\cite{ZGLSN}.
All the results about the transmission and reflection with Fano line and
Breit-Wigner line are valid in the present studies.
To emphasize the role of the dressed states formed by atomic collective
excitations coupled to the field of photon, we still apply the discrete coordinate
equations (\ref{tr-09}) to calculate the transmission coefficient.
\begin{figure}[tbp]
\includegraphics[bb=69 394 498 539, width=7 cm,clip]{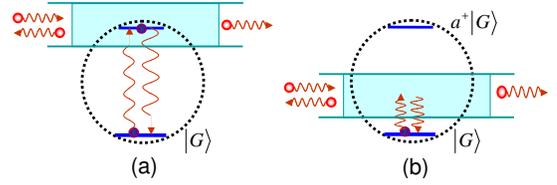}
\caption{(Color online) Diagrammatic presentation for (a) perfect reflection
when $\Omega $ is inside the band; (b) the increasing transmission when $%
\Omega $ is outside the band.}
\label{qnode-4}
\end{figure}
We assume an incoming wave within the CRW with energy $E_{k}$, incident from the left,
results in a reflected and transmitted wave. The wave functions in the
asymptotic regions on the left and right are given by
\end{subequations}
\begin{equation}
u_{j} =\left\{
\begin{array}{c}
e^{ikj}+re^{-ikj}\text{ \ }j\leq -1 \\
se^{ikj}\text{ \ \ \ \ \ \ \ \ }j\geq 1
\end{array}
\right.   \label{tr-08}
\end{equation}
By canceling the probability amplitude $u_{A}$ and $u_{B}$ in the boundary
condition of Eq.(\ref{tr-09}), the transmission amplitude is obtained as
\begin{equation}
s=\frac{-2ig\left( E_{k}-\Omega \right) \sin k}{\left( E_{k}-\Omega _{+}\right)
\left( E_{k}-\Omega _{-}\right) -2ge^{ik}\left( E_{k}-\Omega\right) }\text{,}
\label{tr-10}
\end{equation}
and the reflection amplitude also can be gotten by the relation $r=s-1$.

We notice that the equation (\ref{tr-10}) is the same as that obtained
in Ref.\cite{ZGLSN}. This point can be obviously seen by substituting
the energy dispersive relation $E_{k}=\omega -2g\cos k$ into Eq.(\ref{tr-10}).
However, from the dressed state based representation in Eq.(\ref{tr-10}),
the physical effects of photon scattering process can be feasibly
explained with the two virtual channel pictures mentioned above.
It can be observed from the above equations: i) transmission vanishes at the edge
of band regardless of the location of energy level spacing, and these trivial zeros
are caused by the vanishing group velocity at the $k=0,\pm \pi $; ii) a
vanishing transmission appears, once the energy level spacing $\Omega $ is inside
the energy band $\left[ \omega -2g,\omega +2g\right] $. Case two is related
to the Fano resonance\cite{Fano}, which is the interference effect
characterized by a certain discrete energy state interacting with the
continuum. Indeed, as shown in Fig.\ref{qnode-4}, the atomic ensemble
provides the discrete energy state and the CRW provides the continuum here.
It is the inside-band discrete state shown in Fig.\ref{qnode-4}(a), which
allows additional propagating path for the incident photon, and the
destructive interference leads to perfect reflection.

\begin{figure}[tbp]
\includegraphics[bb=34 282 518 604, width=7.5 cm,clip]{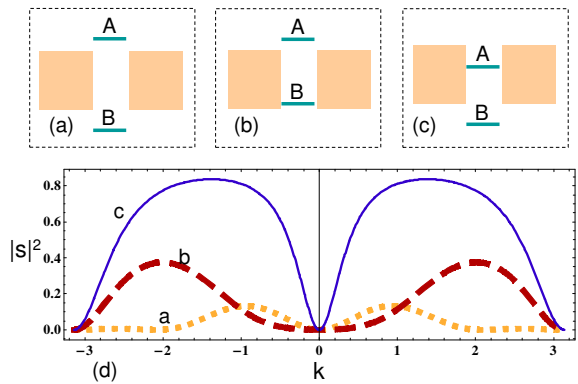}
\caption{(Color online) Energy-level diagram corresponding to the strong
coupling and the transmission coefficient versus the wave number when
$\sqrt{N}\xi =3$.
In: (a) the two energy levels of polaritons are outside the band;
(b) the energy level of polariton $A$ is outside the band and $B$ is inside;
(c) the energy level of polariton $A$ is inside the band and $B$ is outside;
(d) the transmission coefficient, a-line(yellow dot line) at $\omega =3,\Omega =2$,
b-line(red dash line) at $\omega =5,\Omega =8$, c-line(blue solid line)
at $\omega =15,\Omega =5$. Here,
$k$ in units of lattice constant, others are in units of hopping constant $g$.}
\label{qnode-56}
\end{figure}
All the above observations can be clearly seen in Fig.\ref{qnode-56}.
We compare Fig.\ref{qnode-56}(d) three lines, the
transmission probability gets much larger on the whole when the energy level
spacing is tuned outside the band. The physical mechanism for this increasing
transmission is schematically illustrated in Fig.\ref{qnode-4}(b). In the
scattering process when the energy level spacing is outside the band, the
energy $E_{k}$ carried by the incident photon cannot excite quasi-spin-wave
in the atomic ensemble, e.g. it occurs only a virtual exchange process of
photons. However, since the atomic ensemble begins at the initial state
$\left\vert G\right\rangle $ with all atoms in a ground state and almost
ended with the same state $\left\vert G\right\rangle $ during the scattering
process.

Let us understand the above increasing phenomenon of transmission from the point
view of polaritons. The CRW on the left side of the $0$th resonator forms a
continuum with a propagation allowing bound $E_{k}$, and so does the right
side of the $0$th resonator. The two continuum are connected by two discrete
energy states at the point $j=0$. These two discrete energy states
correspond to the single quasi-particle excitation of different polaritons
from the atomic-ensemble ground state $\left\vert G\right\rangle $.
According to the previous definition in Eq.(\ref{tr-04}) and Eq.(\ref{tr-05}),
the required energy for a single excitation of polariton $A$ is higher
than that for a single excitation of polariton $B$. Fig.\ref{qnode-56}(a), (b) ,and(c)
schematically shows the corresponding energy-level diagram.
In Fig.\ref{qnode-56}(a), the two energy levels of polariton are outside
the band, the strong coupling removes both the
single-excitated energy-level $a^{\dag }\left\vert G\right\rangle $ and the
state $b_{0}^{\dag }\left\vert 0\right\rangle $ of the resonator far away
from the band, which almost blocks the tunneling process. Therefore the
transmission is very small on the whole (see the Fig.\ref{qnode-56}(d)'s a-line).
When one of the polariton's energy level is inside the band, there is some
possibility for one discrete energy state inside the band
just like Fig.\ref{qnode-56}(b) and (c), which mediates
the photonic hopping among the resonators. However, if the energy of one
polariton approaches the middle of the band, the transmission probability
becomes higher (see the Fig.\ref{qnode-56}(d)'s b-line and c-line).

\begin{figure}[tbp]
\includegraphics[width=8 cm, height=3 cm]{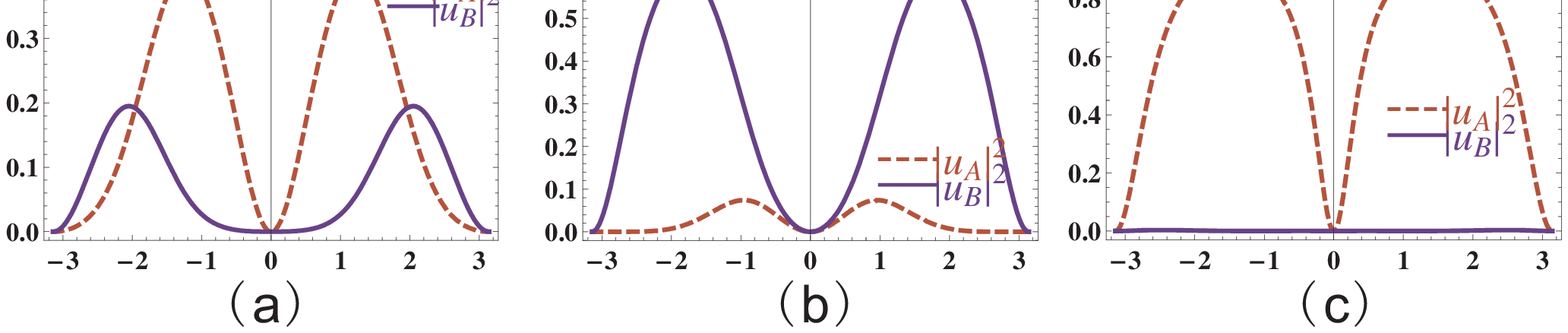}
\caption{(Color online). The probability $|u_{A}|^{2}$ (red dash line) and
$|u_{B}|^{2}$ (blue solid line) versus the wave number $k$, for a
single-photon to form the polaritons A and B. The atomic-cavity coupling
$\sqrt{N}\xi =3$. In: (a) $\omega =3$, $\Omega =2$, (b)
$\omega =5$, $\Omega =8$, (c) $\omega =15$, $\Omega =5$. Here, $k$ in
units of lattice constant, others are in units of hopping constant $g$}
\label{qnode-7}
\end{figure}

Finally let us come to the question which path will be taken with large
probability in the scattering process of the single photon. Actually this
question is answered intuitively in the discussion of the previous
paragraph. We now give the relation between the single-photon scattering
process and the probability for a photon to occur in the polaritons.
From Eq.(\ref{tr-09}), the probability amplitude is obtained as
\begin{subequations}
\begin{eqnarray}
u_{A} &=&\frac{E_{k}-\Omega _{-}}{E_{k}-\Omega }\frac{\xi _{A}}{g}s, \\
u_{B} &=&-\frac{E_{k}-\Omega _{+}}{E_{k}-\Omega }\frac{\xi _{B}}{g}s \text{.}
\end{eqnarray}
\end{subequations}
In Fig.\ref{qnode-7}, we plot the probability $\left\vert u_{A}\right\vert
^{2}$ and $\left\vert u_{B}\right\vert ^{2}$ as the function of wave number
$k$. The read dash line presents the probability $\left\vert u_{A}\right\vert
^{2}$, and the blue solid line presents the probability $\left\vert
u_{B}\right\vert ^{2}$. The parameters are taken as the same as
Fig.\ref{qnode-56}(d). In Fig.\ref{qnode-7}(a), the atomic energy level spacing
$\Omega $ is inside the band. The case for $\Omega $ outside the band is
depicted in Fig.\ref{qnode-7}(b,c).

In Fig.\ref{qnode-7}(b), the atomic energy level
spacing $\Omega $ lies upper the band, while in Fig.\ref{qnode-7}(b), $\Omega$
lies below the band. It can been found that when $\Omega $ is inside the
band, $\left\vert u_{A}\right\vert ^{2}$ and $\left\vert u_{B}\right\vert
^{2}$ compare with each other. However, when the eigen-energy $\Omega _{-}$
of the polariton $B$ is inside the band, as shown in Fig.\ref{qnode-56}(b), there
is a large probability for a single-photon to form the polariton $B$ at the
point $j=0$. Also when the energy level $\Omega _{+}$ of polariton $A$ is
inside the band, the probability $\left\vert u_{A}\right\vert ^{2}$ becomes
much larger, and $\left\vert u_{B}\right\vert ^{2}$ becomes much smaller.
Therefore, photon tends to form polariton $A$ rather than polariton $B$.

The above observation can been understood from the difference between the
coupling strength $\xi _{A}$ and $\xi _{B}$. For the case shown in
Fig.\ref{qnode-7}(a), the coupling strength $\xi _{A}$ and $\xi _{B}$
approximately equal, therefore, $\left\vert u_{A}\right\vert ^{2}$ compares with
$\left\vert u_{B}\right\vert ^{2}$. However, in Fig.\ref{qnode-7}(b,c),
things become different. $\xi _{A}$ is much smaller than $\xi _{B}$ for the
case shown in Fig.\ref{qnode-7}(b), while for the case shown in
Fig.\ref{qnode-7}(c), the coupling strength $\xi _{B}$ is much smaller than
$\xi _{A}$. Therefore, by adjusting the energy level spacing of the atomic
ensemble, one can choose the formation of polaritons in the $0$th cavity.

\section{\label{Sec:boundstate}Localized photons induced by polaritons}

All the above arguments are based on the scattering approach
with the discrete coordinate representation. In contrast
to such delocalized states of scattering, there also exists
the photon localization due to the formation of the polaritons,
which are some quasi-particles described by the bound state.
Actually, the bound state in continuum was first proposed
in 1929 by von Neumann and Wigner\cite{Neumann1929}.
From then on, there are so many studies to report its existence
\cite{Stillinger75,Gazdy77,Capasso92,Ordonez06,Nakamura07,Bulgakov08}.
The bound states are eigenstates of the whole system with some
spatially localized properties. They are locally produced
by lack of the periodicity in coordinator space and their corresponding
eigenenergy is outside the energy band.

To find such bound state in our system, we return to the polariton free
representation of scattering equation.
\begin{equation}
H=\sum_{j}
\omega b_{j}^{\dagger}b_{j}+g\left(  b_{j}^{\dagger}b_{j+1}
+h.c.\right)  +\Omega a^{\dagger}a+\sqrt{N}\xi\left(  b_{0}%
^{\dagger}a+h.c.\right)
\label{bs01}
\end{equation}
The above original Hamiltonian (\ref{bs01}) is obviously equivalent
to Eq. (\ref{tr-06}) with the polariton representation.

We assume
\begin{equation}
\left\vert \Psi\right\rangle _{b}=\sum_{j}u^{b}_{j}\left\vert g,1_{j}%
\right\rangle +u_{e}\left\vert e,0\right\rangle
\end{equation}
is a bound state with the site-dependent amplitude
\begin{equation}
u^{b}_{j}=Be^{-i|k|j}
\end{equation}
Here, $k$ can be regarded the wave vector that connected with the
energy by the dispersion relation for Bloch states.

The eigen-equation
$H\left\vert \Psi\right\rangle _{b}=E_{b}\left\vert \Psi\right\rangle _{b}$
determines a set of discrete coordinator equations
\begin{subequations}
\begin{align}
g\left(u^{b}_{j+1}+u^{b}_{j-1}\right)+\omega u^{b}_{j}+\sqrt{N}\xi u_{e}\delta_{j,0}
  & =E_{b}u^{b}_{j} ,\\
\sqrt{N}\xi u^{b}_{0}+\Omega u_{e}  & =E_{b}u_{e} ,
\end{align}
\end{subequations}
which is the same as Eq.(\ref{tr-09}) with the polariton based representation.

As discovered in Refs.\cite{ZGLSN,gongzr}, the above equations can be reduced
into a scattering equation
\begin{equation}
(E_{b}-\omega)u^{b}_{j}+V_{j}u^{b}_{j}=g\left(u^{b}_{j+1}+u^{b}_{j-1}\right)
\label{bs02}
\end{equation}
with the effective potential
\begin{equation}
V_{j}=-\frac{N\xi^{2}}{E_{b}-\Omega}\delta_{j,0} .
\end{equation}
This local potential depend on the energy $E_{b}$.
It is crucial that this potential is resonant at the energy $E_{b}$
equal to the atomic energy level spacing $\Omega$.
The resonance enhanced coupling to the Zero'th site will lead to a
bound state with the imaginary momentum vector,
so that $u^{b}_{j}$ being exponentially decay with
$j$ is far away from the zero site.

This complex momentum vector $k$ can be determined by the equation
\begin{subequations}
\begin{align}
E_{b}&=\omega+2g\exp(-ik)+\frac{N\xi^{2}}{E_{b}-\Omega} \label{bs03a} ,\\
E_{b}&=\omega+2g\cos k \label{bs03b} ,
\end{align}
\label{bs03}
\end{subequations}
straightly formally given by Eq.(\ref{bs02}). Then we obtain the
transcendental equation for calculating the bound energy as
\begin{equation}
\sin k=\pm\sqrt{1-\left(  \frac{E_{b}-\omega}{2g}\right) ^{2}} .
\end{equation}
The above calculation gives the energy of the bound state as
\begin{equation}
E_{b} =\Omega-\frac{iN\xi^{2}}{2g\sin k} .
\end{equation}
When $E_{b}>\omega_{b}+2g$, and the energy $E\in\mathbb{R}$, we can obtain
only one the bound state energy satisfied all above requests from equation
\begin{equation}
E_{b1}=\Omega+\frac{N\xi^{2}}{\sqrt{(E_{b1}-\omega)^{2}-4g^{2}}}.
\end{equation}
We assume $k=a-ib$, $a\in\mathbb{R}$ and $b>0$.

\begin{figure}[tbp]
\includegraphics[bb=73 286 486 534, width=8 cm,clip]{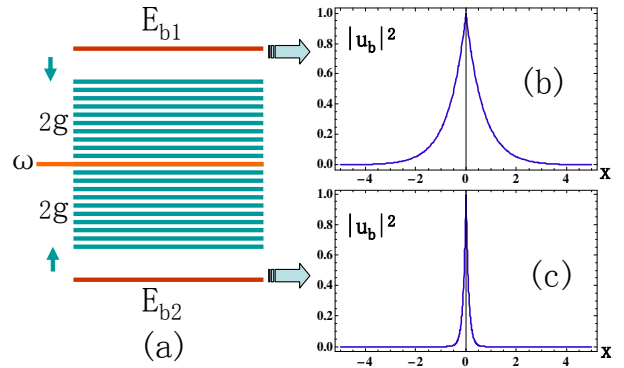}
\caption{(Color online) (a) Diagram of energy spectrum.
$E_{b1}$ and $E_{b2}$ are bound states energy-level diagram in k-space.
They are outside the energy band of wave guide.
Their wave functions show in (b)and(c) correspondingly,
where $\sqrt{N}\xi =3$, $\omega =15$, $\Omega =5$. Here, $j$ in
units of lattice constant, others are in units of hopping constant $g$.}
\label{qnode-9}
\end{figure}

Considering Eq.(\ref{bs03b}) and the energy $E$ is real, we
obtain $a=0$. So we can obtain complex momentum vector $k$,
that means it is one of the bound state. Similarly, when
$E_{b}<\omega_{b}-2g$ we can obtain the another bound state
and it's corresponding energy $E_{b2}$.
We illustrate the bound state in the Fig.\ref{qnode-9}(a)
in the k-space and give the corresponding wave functions
in the Fig.\ref{qnode-9}(b) and Fig.\ref{qnode-9}(c) respectively.
The above investigation display that there exist two localized single
photon states with different energy.

\section{\label{Sec:conclusion}Conclusion with a remarks}

In summary, we have studied the scattering process of photons confined
in a one dimensional optical waveguide by a laser controlled atomic ensemble.
We show the possibility to bound single photon with an atomic
ensemble based quantum node.
Using the discrete coordinate approach, we exactly solve the effective
scattering equations and gave the transmission rate of single photon
in the coupled resonator array.

This investigation motivates us to proposed an active coherent control scheme
for photon transferring in a waveguide by the localized atom ensemble.
This construction can also create a local photon state called bound state.
This finding implies a possibility to store the single photon state locally.

This work is supported by NSFC No.~90203018, No.~10474104, No.~60433050,
No.~10325523, No.~10347128, No.~10075018 and No.~10704023, NFRPC No.~
2006CB921205 and 2005CB724508, and the Scientific Research Fund of Hunan
Provincial Education Department of China (Grant No. 07C579).

\end{document}